\documentstyle[12pt]{article}
\begin{document}
\begin{center}
{\Large \bf Classical many--particle distribution functions:
some new applications

}
\bigskip
{\large E.E.~Tareyeva, V.N.~Ryzhov}
\date{}
\smallskip

{\it
Vereschagin Institute for High Pressure Physics, Russian Academy
of Sciences, 142092 Troitzk,
Russian Federation
}
\end{center}

\begin{abstract}
We present a new purely
equilibrium microscopic approach to the description
of liquid--glass transition in terms of space symmetry breaking
of three-- and four-- particle distribution functions in the
cases of two and three dimensions, respectively.
The approach has some features of spin glass theories as
well as of density--functional theories of freezing.
\end{abstract}

The main purpose of the report is to present a new purely
equilibrium microscopic approach to the description
of liquid--glass transition in terms of space symmetry breaking
of three-- and four-- particle distribution functions in the
cases of two and three dimensions, respectively. The approach
has some features of the spin glass theories as well as of the
density--functional theories (DFT) of freezing.

It is usually believed that there are two essential differencies
between spin glasses and real structural glasses: 1) in the
Hamiltonian of spin glasses there is explicit randomness from
the very beginning, while in the case of real glasses there is
no such randomness. 2)In experiments with spin glasses there is
always the range of the concentration of magnetic impurities
where nothing else that a spin glass phase appears while in the
case of space glass there exists a crystalline ground state.
However, in real systems one can consider these differencies
simply as time scales differencies for the freezing of
corresponding degrees of freedom with respect to the time
scale of the real or computer experiments. In fact, there are
now some indications that two possible candidates for
equilibrium glasses do exist:  some
polydisperce hard--sphere systems and some binary mixtures of
hard spheres. Even if it is not so, it seems us
that one needs an "underlying" equilibrium theory of
liquid--glass transition to understand what really glasses
present as space symmetry breaking problem.
We should mention that beautiful and fruitful time--dependent
mode--coupling theory ~\cite{gotz} which describes a number of
subtle experimental facts does not consider the problem of
space symmetry breaking.  Some other arguments can be found in
the recent papers by Parisi (see, e.g.~\cite{par} and references
therein).

To describe different kinds of space symmetry breaking we use
the formalism of classical many particle conditional
distribution functions
\[ F_{s+1}({\bf r}_1|{\bf r}_1^0 ... {\bf r}_s^0)=
\frac{F_{s+1}({\bf r}_1, {\bf r}_1^0,...,{\bf r}_s^0)}
{F_s({\bf r}_1^0,...,{\bf r}_s^0)}. \]
Here $F_s({\bf r}_1,...,{\bf r}_s)$ is the $s$--particle distribution
function. The functions $F_{s+1}({\bf r}_1|{\bf r}_1^0 ... {\bf
r}_s^0)$ satisfy the equation
\cite{arin}
\begin{eqnarray}
\frac{\rho F_{s+1}({\bf r}_1|{\bf r}_1^0 ... {\bf r}_s^0)}{z}& = &\exp \left\{
-\beta \sum_{k=1}^s \Phi({\bf r}_1-{\bf r}_k^0)
+\sum_{k \geq 1} \frac{\rho^k}{k!} \int \,
S_{k+1}({\bf r}_1,...,{\bf r}_{k+1}) \right. \nonumber \\
& &\left.\times F_{s+1}({\bf r}_2|{\bf r}_1^0 ... {\bf r}_s^0)...
F_{s+1}({\bf r}_{k+1}|{\bf r}_1^0 ... {\bf r}_s^0)
d{\bf r}_2... d{\bf r}_{k+1} \right\} \label{main}.
\end{eqnarray}
Here
$z $ is the activity, $\rho$ is the mean number density, $S_{k+1}({\bf
r}_1,...,{\bf r}_{k+1})$ is the irreducible cluster sum of Mayer functions
connecting (at least doubly) $k+1$ particles,
$\beta=1/k_B T$ and $T$ is the temperature.

If one takes the derivative of
(\ref{main}) relative to ${\bf r}_1$, one obtains the
equilibrium Bogolubov hierarchy \cite{NNB2}
along with the explicit expression for
$F_{s+2}$ as the functional on $F_{s+1}$ which gives
the formally exact closure. However it contains
infinite series and integrals and one has to use some
approximations to exploit it. The same can be said about the
Eq.(\ref{main}) itself.

Let us now consider the symmetry breaking of the one--particle
distribution function and formulate briefly  DFT of freezing
(see ~\cite{dft} and the reviews ~\cite{rev}).
The equation ~(\ref{main}) for $s=0$ is the extremum condition
for the free energy
functional of the inhomogeneous system with the density
$\rho({\bf r}) = \rho F_1(\bf r)$ and has the form:
\begin{equation} \begin{array}{lcl} {\cal F}/k_BT = \int\,d{\bf
r}_1\, \rho({\bf r}_1)[\ln(\lambda^d\rho({\bf r}_1)-1]-\\ -
\sum_{k \geq 1} \frac{1}{(k+1)!} \int \cdots \int\, S_{k+1}({\bf
r}_1...{\bf r}_{k+1})\rho({\bf r}_1) \cdots \rho({\bf
r}_{k+1})\,d{\bf r}_1 \cdots d{\bf r}_{k+1}.\end{array}
\label{free} \end{equation} or \begin{equation} {\cal F}/k_BT =
\int\,d{\bf r}_1\, \rho({\bf r}_1)[\ln(\lambda^d\rho({\bf
r}_1)-1]-{\cal F}_{ex}[\rho({\bf r})]/ k_BT.  \label{frex}
\end{equation} The excess free energy ${\cal F}_{ex}[\rho({\bf
r})]/k_BT$ is just the generating functional for direct
correlation functions \begin{equation} c_n({\bf r}_1...{\bf
r}_n)=\frac{\delta^n {\cal F}_{ex}[\rho({\bf r})]/k_BT} {\delta
\rho({\bf r}_1) \cdots \rho({\bf r}_n)},  \end{equation}
so that Taylor expansion around the liquid can be written
in the following form:
\begin{equation} \beta \Delta F = \int
d{\bf r} \varrho ({\bf r}) \ln \frac {\varrho ({\bf r})}
{\varrho _0} - \sum_{k \geq 2} {1 \over k!} \int c^{(n)} ({\bf
r}_1,...,{\bf r}_k) \Delta \varrho ({\bf r}_1)...\Delta\varrho
({\bf r}_k) d{\bf r}_1 ... d{\bf r}_k , \label{exfree}
\end{equation}
where
$$ \Delta \varrho ({\bf r}) = \varrho ({\bf r}) - \varrho_l $$
is the local density difference between solid and liquid phases.

The full system of equations to be solved in  DFT contains
the nonlinear integral equation for the function $\rho ({\bf
r})$, obtained as the extremum condition for the free
energy and the equilibrium conditions for
 the chemical potential and  the pressure
written in terms of the same functions as
in (\ref{exfree}).
 To proceed constructively in the frame of DFT we must
choose a concrete form of the free energy functional -- a
kind of closure or truncating -- and we must make an ansatz for
the average density of the crystal. The importance of such an
ansatz follows from the fact that we are dealing with a theory
which is equivalent to Gibbs distribution and one has to break
symmetry following the Bogoliubov concept of quasiaverages
\cite{bogol1}.  Now it is necessary to specify the crystal
symmetry (e.g.lattice type) and to locate the freezing
transition for that particular lattice type
\begin{equation} \begin{array}{rcl}
\Delta\rho({\bf r})&=&\rho_l\sum_{{\bf k}}\varphi_{{\bf k}}e^{i{\bf kr}}=
\rho_l\varphi_0+\rho_l\varphi({\bf r}),\\
\varphi_{{\bf k}}&=&\frac{1}{\Delta}\int_{\Delta}\,\frac{\Delta\rho({\bf r})}
{\rho_l}e^{-i{\bf kr}}d{\bf r}.\end{array}     \label{233}
\end{equation}
The sum is over reciprocal lattice vectors and the integral is
taken over the elementary lattice cell
$\Delta$ .$\varphi_{{\bf k}}$ are the order parameters of the
problem. The DFT approach occurs to be very fruitful and was
used to calculate a lot of melting curves for different
systems.

The 3D DFT scenario of freezing is valid for some 2D systems.
However, there is a number of 2D systems which melts through two
continuous phase transition including intermedeate (so called
hexatic) anisotropic liquid phase. The scenario for such a case
of 2D melting is the well known KTNHY
~\cite{kthny} phenomenological scenario. We develop a
microscopic approach to 2D melting ~\cite{ryta,ryta1}
in the spirit of 3D DFT.
Our approach differs from the standard DFT theory of freezing in two
main points:
First, we allow the Fourier coefficients $\rho_{\bf G}({\bf
r})$ of the one-particle distribution function expanded in a
Fourier series in reciprocal-lattice vectors $\{ {\bf G} \}$:  $
\rho({\bf r}) = \sum_{\bf G} \rho_{\bf G}({\bf r}) e^{i{\bf G
r}} $ to fluctuate and to have amplitude and phase.  Second, we
allow the liquid to be anisotropic: we consider as possible the
existence of a phase with constant density but angular dependent
two-particle distribution function
$F_2({\bf r}_1 -{\bf r}_0) \neq g(r_{10})$.

These two points of generalization define the two new order parameters:
the fluctuating $\rho_{\bf G}({\bf r})$ and the Fourier coefficients
characteristic for the broken symmetry of the function
$F_2({\bf r}_1 -{\bf r}_0)$.
Our approach again is based on the Eq.(\ref{main})
but now, considering hexatic phase, we are dealing with the
bifurcation of the solution for the two--particle distribution
function.
The relative spatial distribution of pairs of particles is characterized
by the function
$F_2({\bf r}_1|{\bf r}_0) = F_2({\bf r}_1 -{\bf r}_0)$.
The vector
${\bf r}_1-{\bf r}_0$
defines the direction of the bond between the molecules at the points
${\bf r}_1$ and ${\bf r}_0$. In the ordinary isotropic liquid the
nearest neighbouring of a given molecule (the first coordination
sphere) has a definite local symmetry, which can be characterized by the
set of bond directions. The local structure of the liquid in the
neighbourhood of a molecule at the point ${\bf r}_0'$ is characterized
by the bond directions ${\bf r}'={\bf r}_2 - {\bf r}_0'$. It occurs
that if the point ${\bf r}_0'$ is at sufficiently large distance from
${\bf r}_0$ then there is no correlation between the directions
${\bf r}={\bf r}_1 - {\bf r}_0$ and
${\bf r}'={\bf r}_2 - {\bf r}_0'$.  In this case after the averaging over
the system as a whole the pair distribution function transforms into
the RDF and the equation (\ref{main}) for $s=1$ has the
solution
$F_2({\bf r}_1-{\bf r}_0) = g(|{\bf r}_1-{\bf r}_0|)$,
which corresponds to ordinary isotropic liquid.
When we approach the anisotropic liquid phase the long--ranged
correlations between the bond directions
${\bf r}$ and ${\bf r}'$ do appear and the averaged two--particle
distribution function depends on the bond direction now.

In the vicinity of the transition one can write
\begin{equation}
F_2({\bf r}_1, {\bf r}_0) = g(|{\bf r}_1-{\bf r}_0|) (1+f({\bf r}_1-
{\bf r}_0)) \label{hex}
\end{equation}
where $f({\bf r}_1-{\bf r}_0)$  has the symmetry of the local
neighbourhood of the particle at ${\bf r}_0$.
The bifurcation point is given by the linearized equation
~(\ref{main}) for $s=1$, namely,
\begin{equation}
f({\bf r}_1-{\bf r}_0)=\int \, \Gamma({\bf r}_1, {\bf r}_0, {\bf r}_2)
f({\bf r}_2-{\bf r}_0)\, g(|{\bf r}_2-{\bf r}_0|)d{\bf r}_2 ,
\label{eighteen}
\end{equation} where
\begin{eqnarray}
\Gamma({\bf r}_1, {\bf r}_0, {\bf r}_2)=&&
 \sum_{k \geq 1} \frac{\rho^{k}}{(k-1)!}\, \int\,
 S_{k+1}({\bf r}_1,...,{\bf r}_{k+1})\, \nonumber\\
&&\times g(|{\bf r}_3-{\bf r}_0|)...g(|{\bf r}_{k+1}-{\bf r}_0|)\,
d{\bf r}_3...d{\bf r}_{k+1}.  \label{nineteen}
\end{eqnarray}

At the same time, when one approaches the line defined by the
bifurcation condition, the correlation radius for the orientation
fluctuations of the pair distribution function diverges. This fact
can be shown with the use of the gradient expansion technique in
the case of the equation  (\ref{main}) for $s=3$, if we write the long
range part of the correlator using the principle of vanishing
correlations (\cite{NNB2}) as:
\begin{equation}
F_4({\bf r}_1, ..., {\bf r}_4)=g(|{\bf r}_1 - {\bf r}_2|)
g(|{\bf r}_3 - {\bf r}_4|) (1+  f_4({\bf r}_1, ..., {\bf r}_4))
\label{f4}
\end{equation}
\[ f_4({\bf r}_1, ..., {\bf r}_4)= f_4(r, R, \rho, \varphi_1, \varphi_2).
\] Here $\varphi_1$ is the angle between the vector ${\bf r}={\bf
r}_1-{\bf r}_2$ and the axis ${\bf R}={\bf r}_2-{\bf r}_3$, $\varphi_2$ is
the angle between the vector ${\bf \rho}={\bf r}_3-{\bf r}_4$  and the
same axis.
We have
$f_4(r, R, \rho, \varphi_1, \varphi_2) \rightarrow 0$ when
$R \rightarrow \infty$ .

The microscopic expressions for the elastic moduli and Frank
constant ~\cite{ryta1} enable us to understand on the microscopic
level whether the 2D melting for any given potential is 3D like
or whether it follows the KTHNY scenario.

Let us consider now a possible description of the liquid--glass
transition in terms of space symmetry breaking for three (four)
particle distribution function in 2D (3D) systems. At high
temperature the nearest neighbours of a molecule can take
different relative positions and there is no short--range order
(SRO). At lower temperature a SRO appears which can be of
different kind at different densities (for phase
transitions in liquids see ~\cite{l-l}). The rotation and
the translation of the clusters of prefered symmetry give rise to
the fact that one-particle and two-particle distribution
functions remain isotropic. If a kind of bond orientational order
(BOO) appears the clusters are oriented in similar way and the
two-particle distribution function becomes to be anisotropic
(as in 2D hexatic phase). However, we can imagine another
situation -- freezing of the symmetry axes of the clusters in
different position. The isotropic phase can be considered as
analogous to the paramagnetic phase (of cluster symmetry axes),
the BOO phase -- to the ferromagnetic phase, and the mentioned
freezed phase -- to a spin glass phase.

Let us consider for simlicity a 2D system.
In the vicinity of the transition one can write (in the
superposition approximation for the liquid)
 \begin{equation}
F_3({\bf r}_1| {\bf r}_1^0, {\bf r}_2^0) =
g(|{\bf r}_1-{\bf r}_1^0|) g(|{\bf r}_1-{\bf r}_2^0|)(1+f_3({\bf
r}_1| {\bf r}_1^0, {\bf r}_2^0) \label{gla}
\end{equation}
In 2D case
$f_3({\bf r}_1| {\bf r}_1^0, {\bf r}_2^0)$ depends in fact
on two distances and two angles
 \begin{equation}
f_3({\bf r}_1| {\bf r}_1^0, {\bf r}_2^0) =
f_3(R_0, \phi _0;R_1, \Theta _1 ),
\label{gla1}
\end{equation}
where
$ {\bf R}_0 = {\bf r}_2^0 - {\bf r}_1^0$,
$ {\bf R}_1 = {\bf r}_1 - {\bf r}_1^0$,
$ {\bf R}_2 = {\bf r}_2 - {\bf r}_1^0$ and $\phi _0$ is the
angle of the vector ${\bf R}_0$ with the $z$ axis,
$ \Theta _1$ -- the angle between ${\bf R}_1$ and ${\bf R}_0$
and $ \Theta _2$ -- the angle between ${\bf R}_2$ and ${\bf
R}_0$.

The linearization of ~(\ref{main})  for $s=2$ gives:
\begin{equation}
f_3(R_0, \phi _0;R_1, \Theta _1)=\int \,
\Gamma'(R_0, \phi _0;{\bf r}_2; R_1, \Theta _1)
f_3(R_0, \phi _0;R_2, \Theta _2)
g(|{\bf R}_2-{\bf R}_0|) g(R_2) d{\bf r}_2,
\label{gla2} \end{equation}
where
\begin{eqnarray}
\Gamma'(R_0, \phi _0;{\bf r}_2; R_1, \Theta _1)&=&
 \sum_{k \geq 1}
 \frac{\rho^{k}}{(k-1)!}\, \int\, S_{k+1}({\bf r}_1,...,{\bf
 r}_{k+1})g(|{\bf r}_3-{\bf r}_1^0|)\, \nonumber\\ &\times&
g(|{\bf r}_3-{\bf r}_2^0|)...
g(|{\bf r}_{k+1}-{\bf r}_1^0|)g(|{\bf r}_{k+1}-{\bf r}_2^0|) \,
d{\bf r}_3...d{\bf r}_{k+1}.  \label{gla3}
\end{eqnarray}
There are two kinds of angles entering the equations and two
kinds of order parameters, cosequently. One angle ($\phi
_0$) fixes the position of one pair of particles of the cluster,
and the other ($\Theta _i$) -- the position of the third particle
in the coordinate frame defined by $\phi _0$. The order
parameter connected with $\Theta _i$ is the
generalization of intracluster hexatic parameter
for the case of different coordinate frames. The order
parameter connected with $\phi _0$ is an analogue of
magnetic moment and in glass--like phase one can consider an
Edwards-Anderson parameter $<\cos \phi  _0 (t) \cos \phi
_0(0)>$.  In such a way we come to the concept of a
"conditional" long range order: if we consider two pairs of
particles at infinite distance from one another then there
exists a preferable possibility for the relative position of the
third particle near each pair. The directions of the bonds in
the pairs of particles themselves are subjects to
spin--glass--like order.  In 3D case the rotation of clusters is
given by matrices $D_{lm}^{l'm'}(\vec \omega _{0i})$ so that we
obtain a kind of orientational multipole glass for the clusters.

This work was partially supported by Russian Foundation for
Basic Researches (Grant No. 98-02-16805).

\end{document}